# IN VITRO HISTOMECHANICAL EFFECTS OF ENZYMATIC DEGRADATION IN CAROTID ARTERIES DURING INFLATION TESTS WITH PULSATILE LOADING

OLFA TRABELSI[†]
*Mines Saint-Etienne, University of Lyon, INSERM, U1059 Sainbiose*
*42023 Saint-Etienne France*
*Sorbonne University, University of Technology of Compiègne, CNRS, UMR 7338, BMBI*
*60205 Compiègne, France*
Email: olfa.trabelsi@utc.fr
https://www.researchgate.net/profile/Olfa_Trabelsi

VIRGINIE DUMAS
*University of Lyon, National School of Engineers of Saint-Etienne, LTDS, UMR 5513 CNRS,*
*42100, Saint-Etienne, France*
Email: virginie.dumas@enise.fr

EDOUARD BREYSSE
*Mines Saint-Etienne, University of Lyon, INSERM, U1059 Sainbiose*
*42023 Saint-Etienne France*
Email: breysse.edouard@gmail.com

NORBERT LAROCHE
*University of Lyon, Jean Monnet University, INSERM, U1059 Sainbiose*
*42023 Saint-Etienne France*
Email: norbert.laroche@univ-st-etienne.fr

STEPHANE AVRIL
*Mines Saint-Etienne, University of Lyon, INSERM, U1059 Sainbiose*
*42023 Saint-Etienne France*
Email: avril@emse.fr

In this paper, the objective is to assess the histomechanical effects of collagen proteolysis in arteries under loading conditions reproducing *in vivo* environment. Thirteen segments of common porcine carotid arteries (8 proximal and 5 distal) were immersed in a bath of bacterial collagenase and tested with a pulsatile tension/inflation machine. Diameter, pressure and axial load were monitored throughout the tests and used to derive the stress-stretch curves and to determine the secant circumferential stiffness. Results were analyzed separately for proximal and distal segments, before and after 1, 2 and 3 hours of enzymatic degradation. A histological analysis was performed to relate the arterial microstructure to its mechanical behavior under collagen proteolysis. Control (before enzymatic degradation) and treated populations (after 1, 2 or 3 hours of enzymatic degradation) were found statistically incomparable, and histology confirmed the alteration of the fibrous structure of collagen bundles induced by the collagenase treatment. A decrease of the secant circumferential stiffness of the arterial wall was noticed mostly at the beginning of the treatment, and was less pronounced after 1 hour. These results constitute an important set of enzymatically damaged arteries that can be used to validate biomechanical computational models correlating structure and properties of blood vessels.

# 1. Introduction

It is known that aortic aneurysms are the result of proteolytic degradation of elastin and collagen in the aortic wall leading to a dilatation of the artery. The enzymes responsible for the degradation of elastin and collagen are called matrix metalloproteases (MMP). They are naturally present in the extracellular matrix (ECM) and can be produced by smooth muscle cells. Several studies showed that the concentration and activity of MMPs are increased in the case of aortic aneurysms [1][2][3] which results in excess ECM destruction and progressive weakening of the arterial wall. Several animal models of aortic aneurysm exist to help investigators to study these mechanisms of disease [4]. For instance, the elastase perfusion model in rats offers aneurysm formation reproducibility and has similar chronic inflammatory infiltrate as seen in humans [5].

Several MMPs are implicated in the aneurysm disease [6], [7], and *in vitro* studies are required to decipher the mechanisms by which proteolytic degradation may induce biomechanical alterations in blood vessels. Biomechanical studies invariably show that the mechanical stability of the arterial wall essentially relies on fibrillar collagens in media and adventitia. These structural collagens are highly resistant toward proteolytic degradation, and the only proteases that have been shown to cleave the native triple helical region of fibrillar collagen, by cutting the bond between a neutral amino acid (X) and glycine in the Pro-X-Glyc-Pro sequence, are the classic collagenases MMP-1, -8, and -13. Single-molecule magnetic tweezers assay have been developed to study the effect of force on collagen proteolysis by MMP-1 [8]. Chang et al. [9] explained these discrepancies and proposed a molecular mechanism by which mechanical force might change the rate of collagen cleavage.

An abundant literature has focused also on the mechanical effects of collagenolysis at the tissue scale. Pioneering work of Roach [10] and then from Dobrin [11] on canine and human vessels treated with collagenase, showed that wall integrity depends on intact collagen rather than elastin. Dobrin [12] also showed that treatment with collagenase reduced the circumferential stress when the vessels were distended by at least 60 mmHg, but it did not reduce the longitudinal stress. Dadgar [13] showed that the action of collagenase induced a loss of stiffness which depended on the duration of exposure. More recently, Gundiah et al. [14] showed that collagenase treated tissues were more compliant in the longitudinal direction as compared to control tissues. Collagenase treatment also caused a decrease in the tissue nonlinearity as compared to the control samples in the study. Gundiah et al. [14] also studied the effect of proteolytic treatments on the tissue thickness. They showed that the samples treated by collagenase were significantly less thick than the control ones, whereas the samples submitted to elastase were thicker than the control ones. Beenakker et al. [15] related the alterations of material properties due to collagenase treatment to the changes of microstructure using multiphoton microscopy.

The aim of all these studies, starting from Dobrin's work, was to determine the roles of collagen and elastin in the biomechanical response of arterial walls. Even very recently, Schriefl et al. [16] tested enzymatic approaches for the selective digestion of collagen and elastin in human AAA. They confirmed that elastin is responsible for the initial stiffness in elastic arteries whereas collagen dominates for large stresses close to rupture.

However, arteries *in vivo* work usually in the intermediate domain where both collagen and elastin contribute significantly to the response. Only few studies considered the effect of enzymatic degradation in this intermediate domain of loading. Gaul et al. [17] identified, for the first time, the strain dependent degradation behavior of arterial tissue using a combination of experimental and theoretical methods. They also showed distinctly different degradation responses when using crude or purified collagenase, highlighting the need to carefully select the appropriate enzyme when investigating tissue degradation, particularly in tests involving highly heterogeneous tissues.

By examining the mechanical properties of tissue strips obtained before and after treatments with elastase and collagenase, even at different strains, one may miss the effects of typical arterial loads which combine circumferential strains (resulting from inflation) and axial strains (resulting from axial tension). Ghazanfari [18] demonstrated, for the first time, that biaxial strain in combination with collagenase alters the collagen fiber alignment from an initially isotropic distribution to an anisotropic distribution with a mean alignment corresponding with the strain at the minimum degradation rate, which may be in-between the principal strain directions. Therefore, tests should reproduce as much as possible the *in vivo* biaxial loading conditions. For instance, Martinez et al. [19] investigated the effects of collagen degradation on the critical buckling pressure of arteries, showing that it was lower after collagenase treatment.

Another important aspect of *in vivo* biaxial loading conditions is the pulsatility, which may play a major role in MMP activation and action on stressed matrix fibers. In the current paper, we study for the first time enzymatic degradation of arteries maintained at their *reference* axial stretch and subjected to cyclic pressures. The objective is to assess the histomechanical effects of collagenase treatment in this specific loading condition that closely reproduces the *in vivo* condition. 3D inflation tests combine both the effect of structural changes (diameter increase, thickness reduction) and material changes (microstructural damage, softening). Previous studies have characterized material effects but not how the structural and material effects are coupled together. For instance, in our study, the result of collagen damage is a larger diameter for the 80-140mmHg range. This larger diameter combined with smaller thickness induce a significant increase of the circumferential stress. Collagenous tissues become stiffer when the circumferential stress increases (structural stiffening), but simultaneously collagenase induces a loss of material stiffness (material softening). Our experiments combine these both effects and look at how they manifest for two specific locations of the carotid artery.

We do not seek to study a particular *in vitro* model of arterial disease, but rather to use collagenase treatment to create a diverse set of enzymatically damaged arteries that may aid in the development of computational models correlating structure and properties of blood vessels.

## 2. Materials and Methods

Tension/inflation tests were carried out on a cohort of 13 porcine carotid samples to better understand their mechanical properties related mainly to collagen enzymatic degradation. Then, a histological study was performed in order to relate the arterial microstructure to its mechanical behavior under the effects of collagen proteolysis.

### 2.1. Sample Preparation

Common left and right carotid arteries were ordered from the Veterinary Campus of the University of Lyon (VetAgro Sup, Marcy l'Etoile) (see Fig.1). Youna castrated pigs, aged between 5 and 6 months and weighing between 80 and 90kg were sacrificed in accordance with the recommendations of the VetAgro Sup Ethics Committee (C2EA No. 18), and in accordance with the regulations on animal experimentation - Directive 2010/63 / EU.

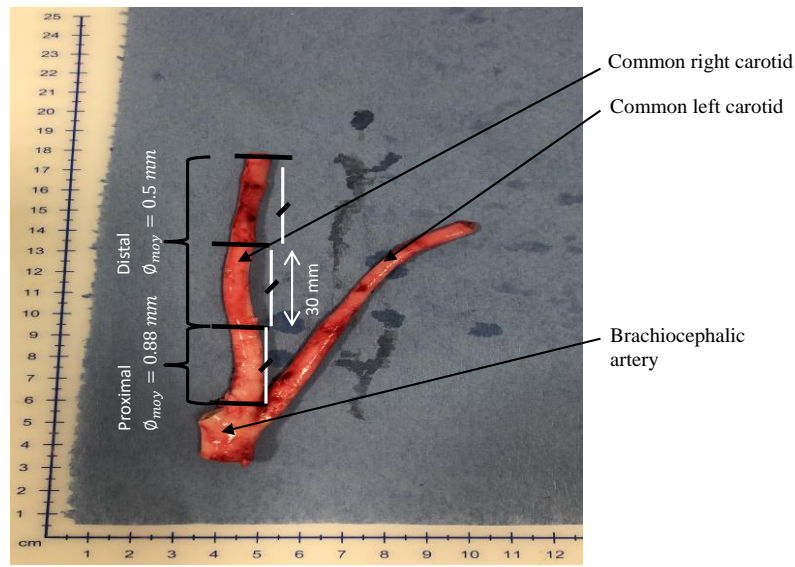

Fig.1: Samples of left and right common carotid arteries.

The samples were stored at −24°C. The postmortem autolysis was minimized by an immediate freezing of the tissue and by performing the tests soon after the samples were thawed.

As the wall thickness and luminal diameter vary from the proximal to the distal position [24], [25], samples were divided in two groups; distal and proximal (see Fig.1).

The wall thickness was measured on all proximal samples and at several points of each specimen. Average proximal thickness value was found to be 0.8 mm. Likewise, the average distal thickness value was found to be 0.5 mm.

### 2.2. Collagenase Treatment

The collagenase was supplied by the company Worthington Biochemical Corporation®. It was a type 2 bacterial (Clostridium histoliticum) collagenase, named "CLS-2", containing MMP1, MMP8 and MMP13. Such collagenase combines collagenolytic and proteolytic activities, which are effective at breaking down extracellular matrices. One component of the substance is a hydrolytic enzyme, which degrades the helical regions in native collagen preferentially at the Y-Gly bond in the sequence Pro-Y-Gly-Pro, where Y is most frequently a neutral amino acid.

The collagenase was supplied in the form of a powder containing 360 U/mg of collagenase. The powder was stored at 4ºC and in a dry place. For the tests, we diluted this powder in PBS to obtain a concentration of 0.56 mg/mL of collagenase, activated at 37°C.

## *2.3. Biomechanical Test*

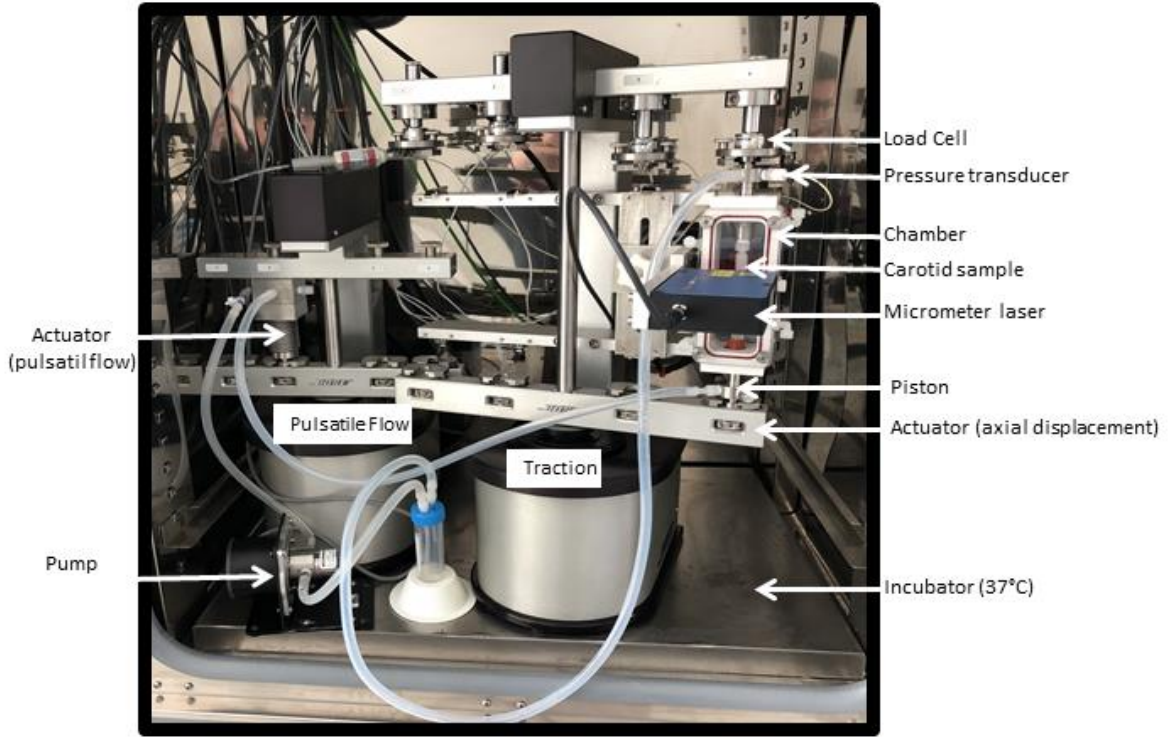

Fig.2: Pulsatile tension/inflation loading for carotid specimens (ElectroForce® BioDynamic® 5270).

Tension/inflation tests were achieved using the bioreactor Biodynamics 5200-Bose (IVTV Platform, ANR-10-EQPX-06-01, FR). Thirty millimeters long segments of carotid arteries were cut, trying to minimize diameter variation in each sample. They were mounted and fixed with sutures. Then, specimens were placed into a chamber, which was then mounted on the Pulsatile Loading Bose bioreactor (ElectroForce® multi-specimen BioDynamic® 5270) to perform tensile/inflation tests. This bioreactor was placed in an incubator at 37°C. Chamber and external circuit were then filled with PBS and the pressure was regulated to 5mmHg.

Diameter variations of the carotid arteries were measured during the tension/inflation tests using a laser based measurement system according to shadow principle, OptoCONTROL ODC2520-46 Laser micrometer (®Micro-Epsilon), with a measurement range of 46 mm and a measurement frequency of 2.5 kHz.

Axial force versus pressure curves were plotted for different values of axial stretch ranging between 1.3 and 1.9, with increments of 0.1. For each axial stretch, the pressure was varied between 5 and 150 mmHg. The axial stretch yielding the least variations of the axial force for an increasing pressure was retained as the preferred *reference* axial stretch $\lambda_{iv}$ [26].

Then the artery was maintained at $\lambda_{iv}$ and inflation/deflation cycles were applied between 30 to 150mmHg at a frequency of 1Hz and with a sinusoidal temporal variation.

The perfusion liquid and the immersion bath both contained the collagenase solution, permitting a progressive enzymatic degradation of the artery under pulsatile loading. At time zero and then every hour, pressure (P) vs diameter (ϕ) curves were plotted for further analysis. We defined the average circumferential stress of the artery according to the Laplace law (1):

$$\sigma_{\theta\theta} = \frac{P\phi}{2h} \qquad (1)$$

Where h is the current thickness, $h_0$ is the initial (loadfree) thickness, and $\lambda_{\theta\theta} = \phi/\phi_0$ is the circumferential stretch.

For the sake of consistency with our radius to thickness ratios, the thick-wall expression was considered, and written as:

$$h = \frac{\phi}{2} - \frac{1}{2}\sqrt{\frac{\phi^2}{2} - 2\frac{2h_0^2 - \phi_0 h_0}{\lambda_{iv}}}$$

This permitted to plot curves of the average circumferential stress for the stress values, and the average circumferential stretch for the stretch values at different stages of degradation of the artery.

The stress / stretch curves were plotted with the wall-averaged values of both the stress and the stretch, as both the circumferential stress and stretches may have variations across the thickness. Variations of wall thickness and diameter along the wall vessel were neglected. The stress / stretch curves were only plotted for the mid-length cross section at which the diameter and the thickness were measured.

Eventually, we defined the secant circumferential stiffness $C_{tt}$ as the average derivative of the circumferential stress with respect to the circumferential strain. The minimum and maximum stress values bounding the domain across which this average was derived corresponded to a minimal pressure of 80 mmHg and a maximum pressure of 140 mmHg. The circumferential strain was defined as the current $\lambda_{\theta\theta}$ divided by $\lambda_{\theta\theta}$ for the diastolic pressure (P = 80mmHg).

*2.4. Histology*

Control and three-hour degraded carotid specimens, were fixed in Ethanol 80% at 4°C and routinely processed for paraffin embedding and cross-sectioned to obtain 5 μm-thick sections (Microtome Leica RM2245, Feather Microtome blade N35HR). Before use, sections were deparaffinized, rehydrated and processed for histochemical staining. The slides were incubated with a 1% Sirius Red solution dissolved in aqueous saturated picric acid for 30 minutes, washed in acidified water (0.5% hydrogen chloride), dehydrated and mounted with Entellan® medium. Collagen type I and III were red stained.

We also performed collagen type I (Coll1α1) immunolabeling. The specimens were incubated at 37°C for 2 hours with an anti-collagen I rabbit primary antibody diluted 1/100 (Abcam, Cambridge, UK, ref. ab292). Then specimens were rinsed in PBS and incubated for 45 minutes with an Alexa-Fluor®-488 conjugated goat anti-rabbit IgG diluted 1/250 (Invitrogen, Life Technologies, Eugene, OR, USA, ref. A011034). The immunolabelling method was chosen to reveal the collagen I fibers and to permit qualitative investigation of their morphology.

After rinsing, specimens were observed with the Zeiss AxioObserver® optical fluorescent videomicroscope and images were obtained with the AxioCam® camera using the AxioVision® software.

*2.5. Statistical Analysis*

A statistical analysis was performed on the secant circumferential stiffness values ($C_{tt}$). Eight groups were considered: groups of proximal and distal common carotid arteries at 0, 1, 2 and 3 hours of enzymatic degradation. For each group, we had n≥5 (Tab 1 and 2) but for the group of distal carotid arteries after 3 hours of enzymatic degradation. The normal distribution was tested by the Anderson–Darling test for the parametric inference (Fig.3).

Tukey's method was used in ANOVA to create confidence intervals for all pairwise differences between mean values while controlling the family error rate to 95%, called Tukey Simultaneous Confidence Intervals.

# 3. Results

## 3.1. Biomechanical Tests

Several tests of pressurization and depressurization were carried out at different axial stretches (from 1.3 to 1.9). Axial force versus inflation pressure curves plotted for these values of axial stretch, showed minimal variations of the axial force for an axial stretch of 1.6. At this axial stretch, the axial force remained almost stable during pressurization and depressurization. For lower axial stretches, the axial force decreased when the pressure increased. For larger axial stretches the axial force increased when the pressure increased. This value was taken as the reference in vivo axial stretch $\lambda_{iv}$. It has been shown previously that the reference axial stretch of carotid arteries is close to the *in vivo* axial stretch [26]. All the samples were axially stretched at 1.6 in the collagenolysis experiments. Samples used to determine the axial stretch could not be used for the other tests as they were damaged when we stretched them beyond 1.6 for determining the *in vivo* axial stretch.

At the reference axial stretch $\lambda_{iv}$, and for a pressure varying from 30 to 150 mmHg, pressure/diameter values were registered during several cycles of pressurization and depressurization. To guarantee the stability of the acquired data, only the tenth cycle was kept. Stress-stretch curves of proximal and distal common carotid arteries before and after one, two and three hours of enzymatic degradation were plotted and analysed (see Fig.3 and Fig.4). All curves for control arteries showed a J-shape with a marked hysteresis. For the proximal control samples, stretches ranged between 1 and 1.27, whereas stresses ranged between 5 and 186 kPa. For the distal samples, stretches ranged between 1 and 1.41, and stresses ranged between 0 and 294 kPa. The stress was found larger in the distal segment as its thickness is lower than the one of the proximal segment.

Generally, in both proximal and distal samples, the collagenase treatment generated a loss of hysteresis (loss of viscoelasticity due to collagen proteolysis), and a decrease of the stress levels for the same value of stretch (loss of stiffness).

In some samples, we observed non-uniform circumferential increase of the diameter, with "blister-like" dilations. Due to these effects, some curves had to be eliminated, especially when these effects disturbed diameter measurements (curve corresponding to 2h collagenase treated specimen in Fig. 3, and curve corresponding to 3h collagenase treated sample in Fig 6).

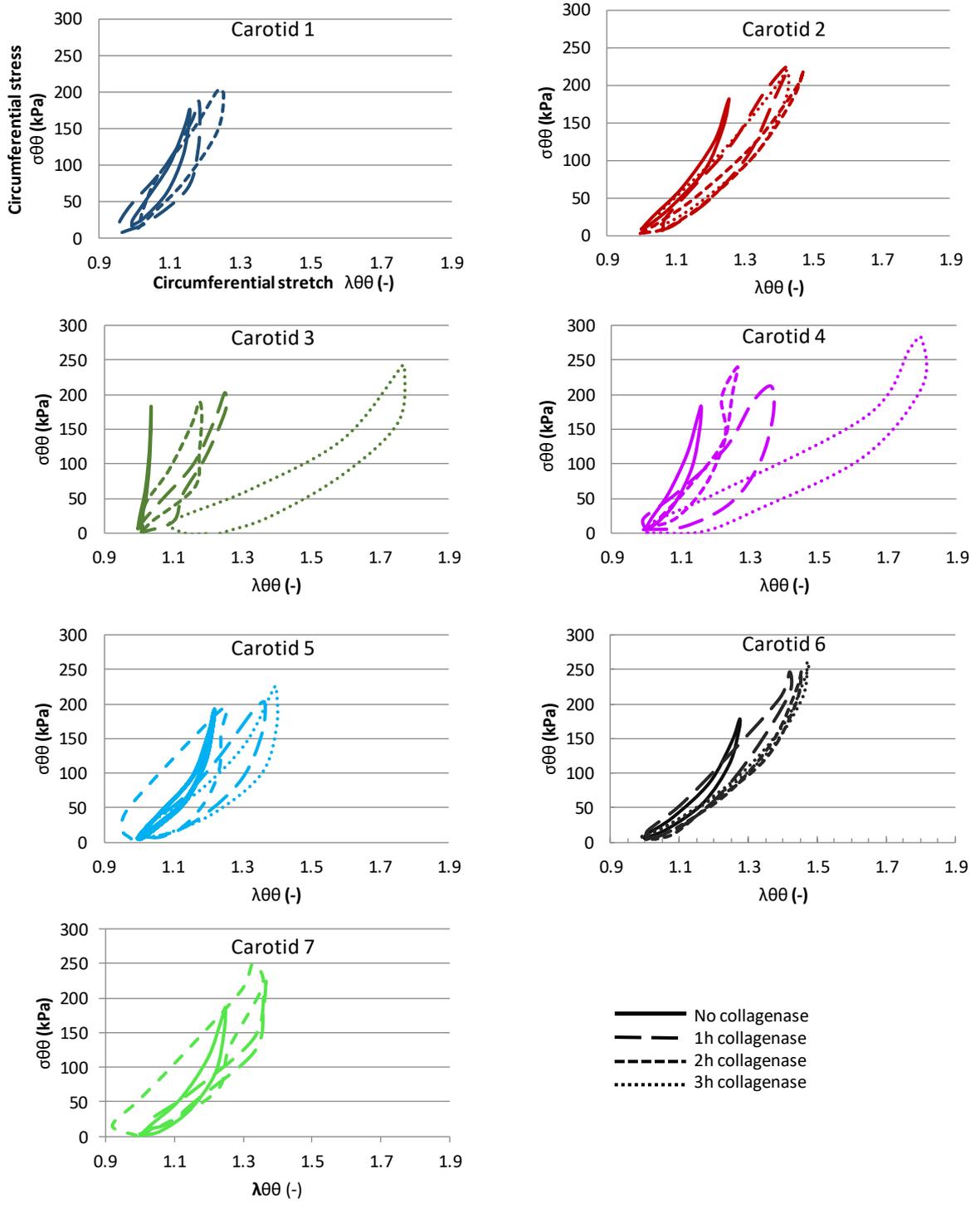

Fig. 3. Stress-stretch curves of 7 proximal common carotid arteries before enzymatic degradation *(solid lines)* and after enzymatic degradation (1h, 2h or 3h) *(straight lines)*.

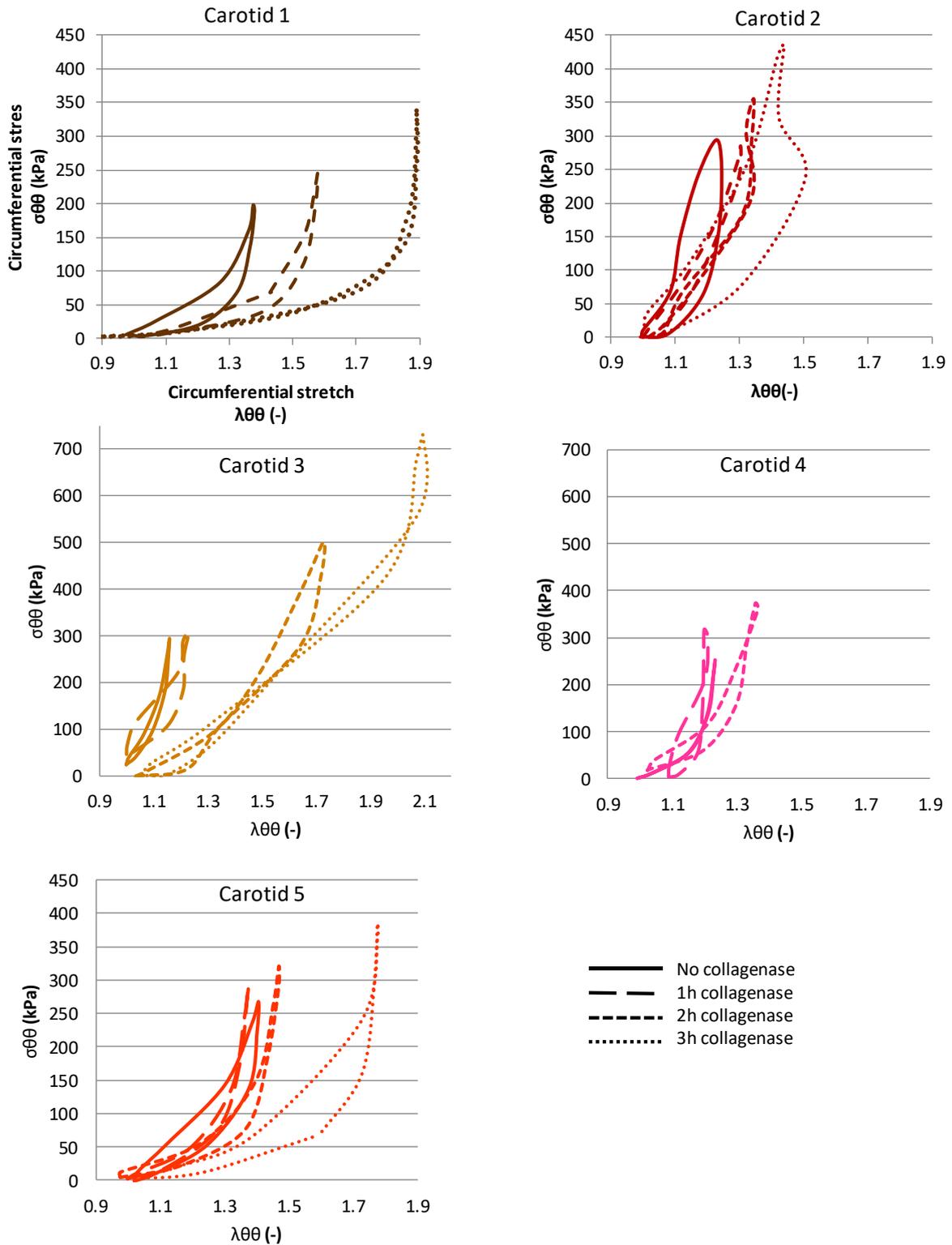

Fig.4. Stress-stretch curves of 5 distal common carotid arteries before enzymatic degradation *(solid lines)* and after enzymatic degradation (1h, 2h or 3h) *(straight lines)*.

The evolution of arterial diameters (distal and proximal carotids) during collagenase degradation at 80mmHg and 140mmHg is shown in Fig.5. It can be noticed that, for a given pressure, the diameter of the carotid increases with enzymatic degradation. This result is also related to Fig.3: for a given stress, the artery reaches a larger stretch (loss of stiffness).

A permanent deformation could also be noticed: arteries showed a diameter increase during collagenase treatment and after every inflation/deflation cycle. The average diameter of the proximal samples before degradation, and at 80 mmHg, was found equal to 7.12 mm. This value increased during collagen degradation to reach 8.01 mm after 3 hours degradation (an increase of 12.5 %). At 140 mmHg the diameter values presented an increase of 17.5% (7.5 mm for the mean value of control diameters and 8.8 mm for the mean value of diameters after 3 hours degradation).

For the distal samples at 80 mmHg, diameter increased from 5.9 mm (mean value for control samples) to 6.83 mm after 3 hours degradation (mean value of diameters after 3 hours degradation). An increase of 15.6% was obtained. Similarly, at 140 mmHg, the diameter values presented an increase of 16.6% (6.15 mm for the mean value of control diameters and 7.36 mm after 3 hours degradation).

The evolution of the secant circumferential stiffness after collagenase degradation was derived from the stress / stretch curves and plotted in Fig. 6. In both distal and proximal carotid samples, a decrease of the secant circumferential stiffness was observed; this decrease being significant after the beginning of the treatment (between 0 and 1 hour). After 1 hour the loss of stiffness was less pronounced.

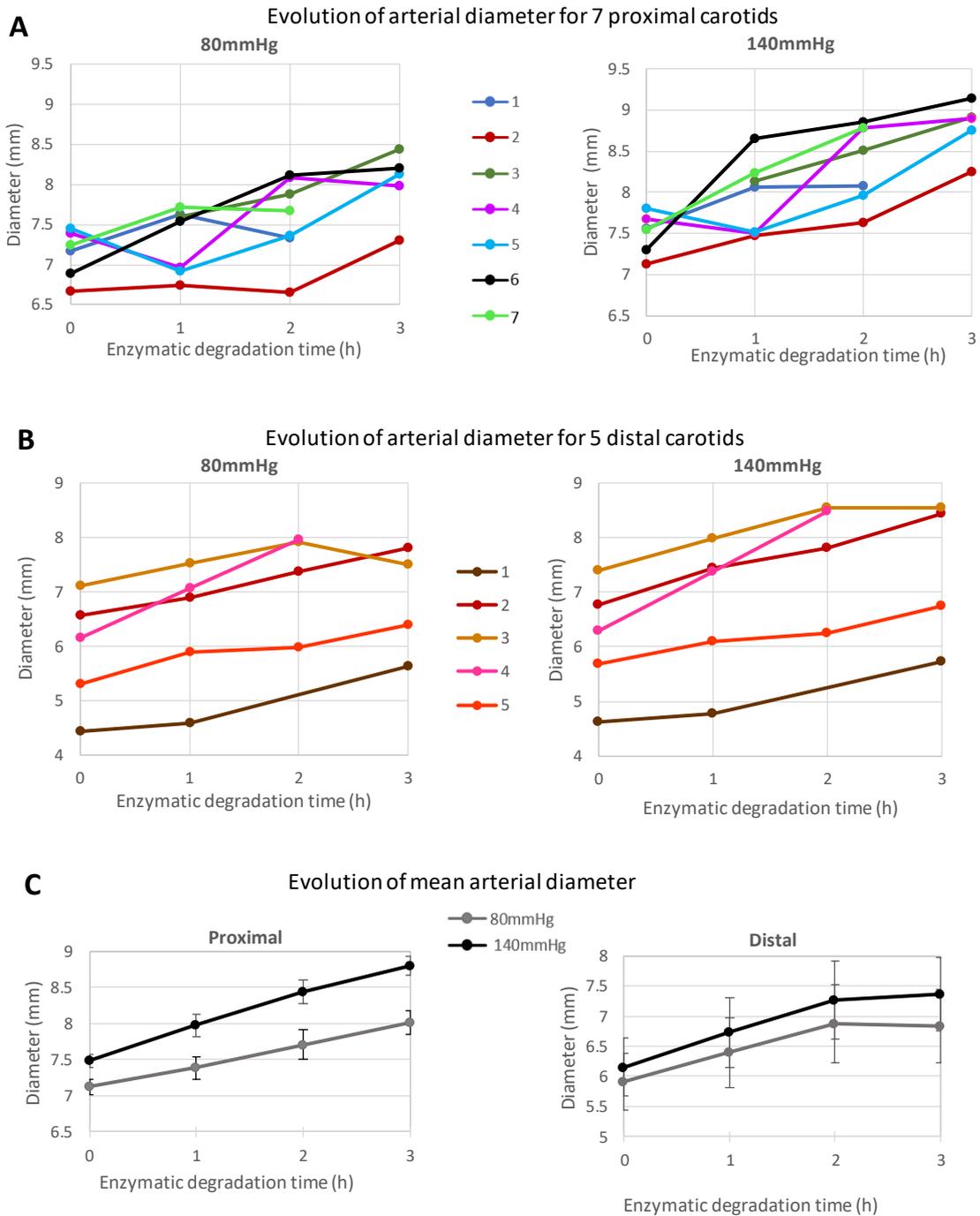

Fig. 5. (A) Evolution of arterial diameter for 7 proximal carotids during collagenase degradation at 80mmHg and 140mmHg. (B) Evolution of arterial diameter for 5 distal carotids during collagenase degradation at 80mmHg and 140mmHg. (C) Evolution of the mean arterial diameter for proximal and distal carotid during collagenase degradation at 80mmHg and 140mmHg (Mean ±SEM n=3 to 7).

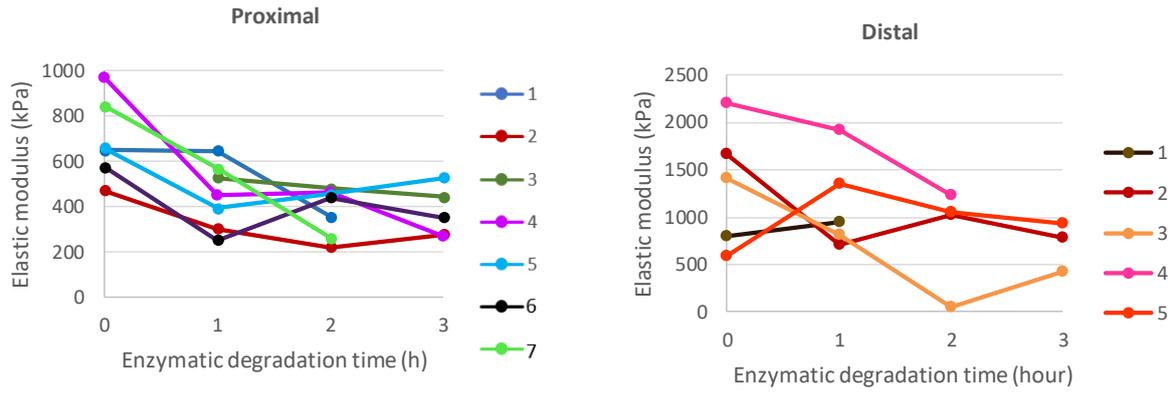

Fig. 6. Evolution of secant circumferential stiffness for 7 proximal carotids and 5 distal carotids after collagenase degradation after 0, 1, 2 and 3 hours.

## 3.2. Histological Analysis

In order to relate the mechanical behavior to the microstructure, a histological analysis was performed. In Fig. 7, histology images for proximal and distal carotid samples, using Sirius Red staining and Immuno-fluorescence of collagen type I, showed an alteration of the fibrous structure of collagen bundles under collagenase treatment. More specifically, after digestion, the fraction of intact fibers notably decreased.

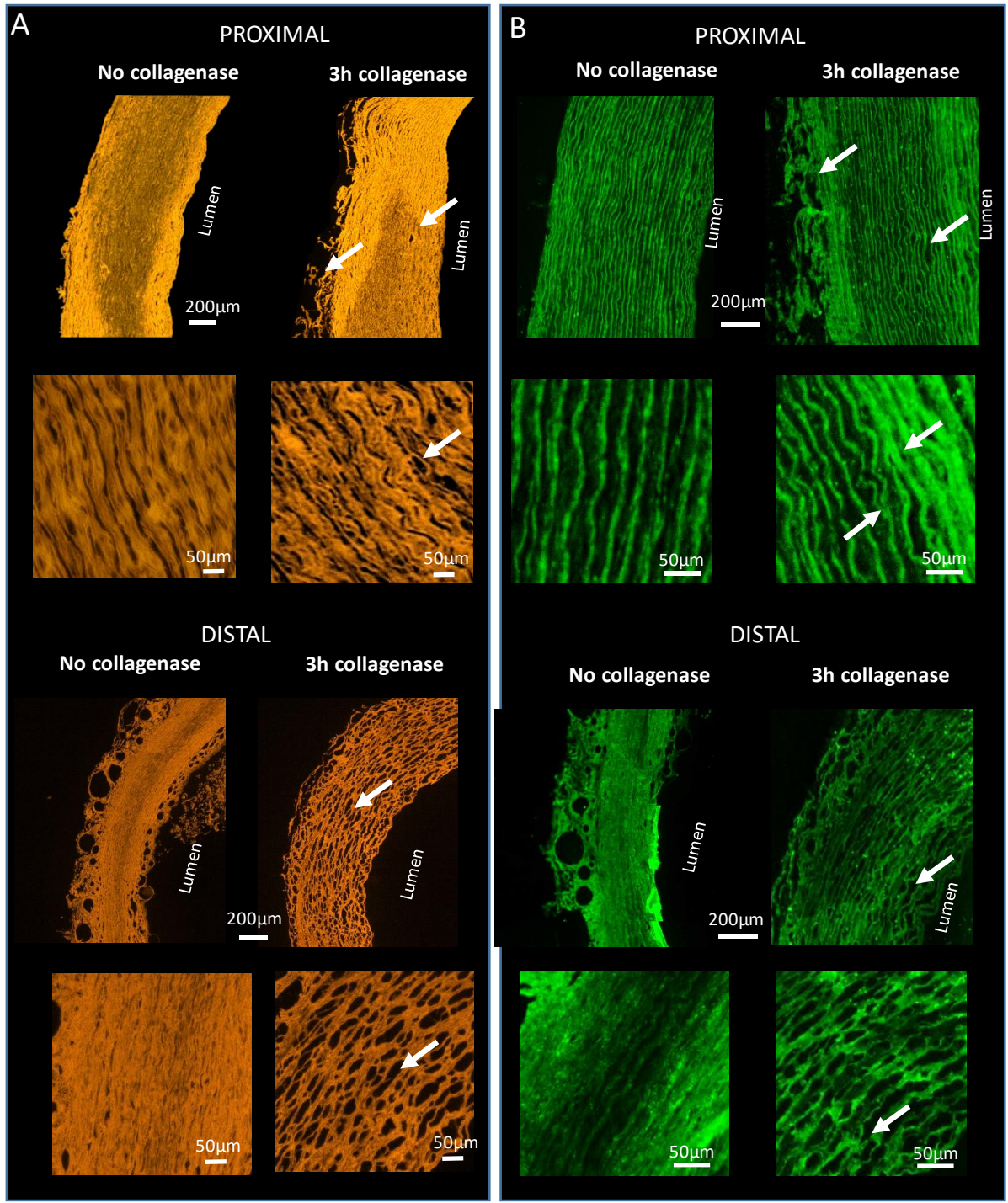

Fig.7. Histological sections of proximal and distal carotids illustrating the degradation of collagen fibers after enzymatic degradation. (A) Sirius red-stained sections (collagens type I, II and III). (B) Immuno-fluorescence analysis of collagen type I. White arrows indicate collagen fibers with a loss of the fibrous structure

### 3.3. Statistical Results

Normal distributions of physiological elastic moduli were verified (p-value ≥0.05) for all groups (control and degraded samples, proximal and distal carotid segments) (Fig.8.). Mean and standard error of mean values are reported in Table 1 and Table 2.

The statistical comparison between each group is shown in Fig.8.

Table 1. Physiological elastic modulus in control and degraded proximal porcine carotid samples

| Proximal carotid | | | | |
|---|---|---|---|---|
| Sample N° | Control carotid | Degraded carotid | | |
| | | 1 hour | 2 hours | 3 hours |
| 1 | 650.06 | 646.96 | 354.07 | - |
| 2 | 472.61 | 303.28 | 221.24 | 276.20 |
| 3 | - | 527.34 | 480.87 | 442.30 |
| 4 | 974.07 | 451.67 | 463.19 | 266.76 |
| 5 | 659.21 | 391.51 | 454.67 | 528.86 |
| 6 | 754.69 | * 1341.73 | *1136.49 | - |
| 7 | 573.80 | 251.01 | 440.03 | 353.28 |
| 8 | 844.44 | 568.57 | 261.76 | - |
| N | 7 | 8 | 8 | 5 |
| P | 0.91 | 0.93 | 0.12 | 0.52 |
| Mean | 704.13 | 448.62 | 382.26 | 373.48 |
| SEM | 63.74 | 54.20 | 39.67 | 45.71 |

(-) Unavailable values
(*) Outlier values

Table 2. Physiological elastic modulus in control and degraded distal porcine carotid samples

| Distal carotid | | | | |
|---|---|---|---|---|
| Sample N° | Control carotid | Degraded carotid | | |
| | | 1 hour | 2 hours | 3 hours |
| 9 | 799.20 | 950.86 | * 6457.51 | - |
| 10 | 1665.52 | 702.54 | 1015.74 | 786.65 |
| 11 | 1408.65 | 806.00 | 47.68 | 428.49 |
| 12 | 2203.69 | 1926.00 | 1224.59 | - |
| 13 | 591.75 | 1354.83 | 1057.22 | 933.91 |
| N | 5 | 5 | 5 | 3 |
| P | 0.77 | 0.34 | 0.06 | 0.41 |
| Mean | 1333.80 | 1148.00 | 836.31 | 716.35 |
| SEM | 292.29 | 223.88 | 266.73 | 150.08 |

(-) Unavailable values
(*) Outlier values

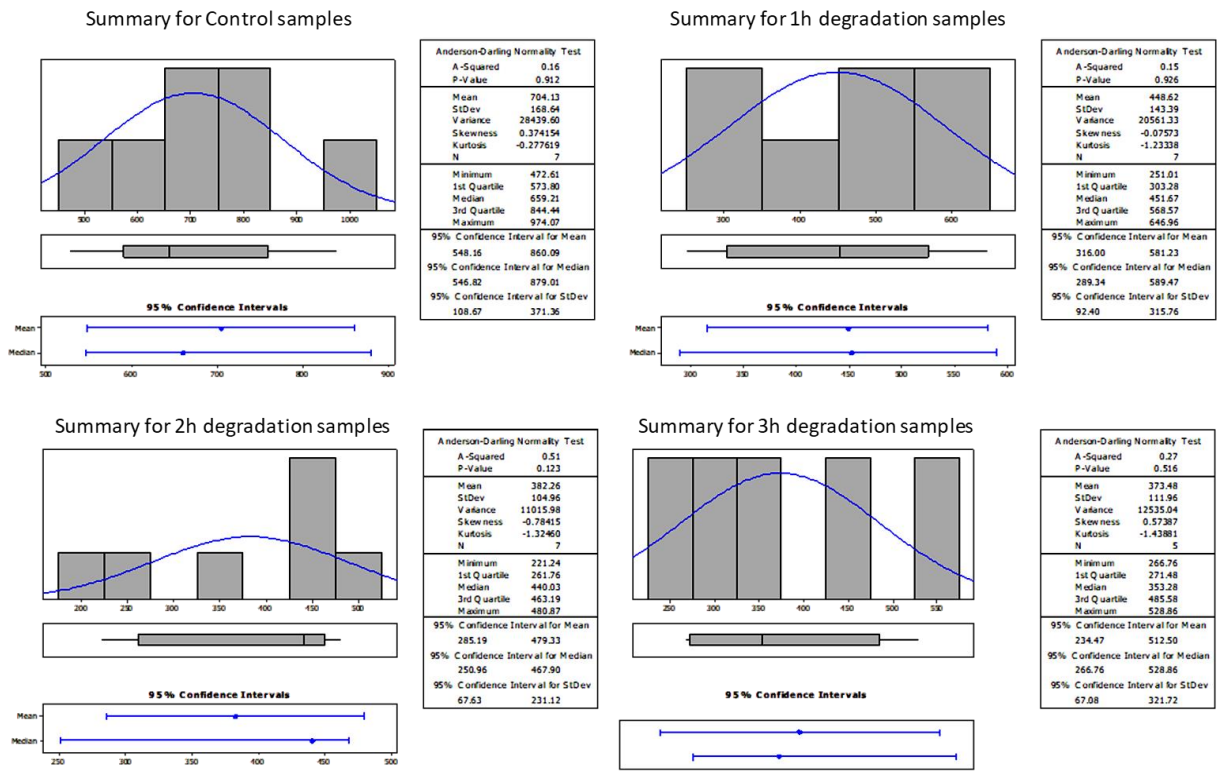

Fig. 8. Statistical verification of the normal distribution of the secant circumferential stiffness in the proximal and distal common carotid segments after 0, 1, 2 and 3 hours of collagenase degradation, using the Anderson–Darling test for the parametric inference.

A significant difference was found in the values of the secant circumferential stiffness between distal and proximal control groups (P=0.033, Individual confidence level "ICL"= 95.00%). This statistical significance supports the choice of dividing samples in two groups (proximal with $h_0$=0.8mm, and distal with $h_0$=0.5mm).

A statistical significance was found with P=0.005 (ICL= 95.00%) between distal and proximal samples degraded with collagenase during one hour. Similarly, a significant difference was found between distal and proximal groups after three hours of degradation (P=0.037, ICL= 95.00%). However, no statistical significance was found between distal and proximal groups after 2 hours of degradation, showing a very low P value (P=0.05, ICL= 95.00%).

For proximal carotid segments, a significant statistical difference was found between the control group (before enzymatic degradation) and all degraded groups (1h, 2h and 3h of degradation) (P=0.001, Individual confidence level = 98.91%). Conversely, no significant difference was found between each of the degraded groups (P=0.497, ICL = 97.99%).

For distal carotids, no statistical significance was found neither between control and 1, 2 and 3 hours degraded samples (P= 0.372, ICL = 98.84%), nor between all degraded samples (P= 0.431, ICL = 97.91%). All these results are summarized in Fig.9.

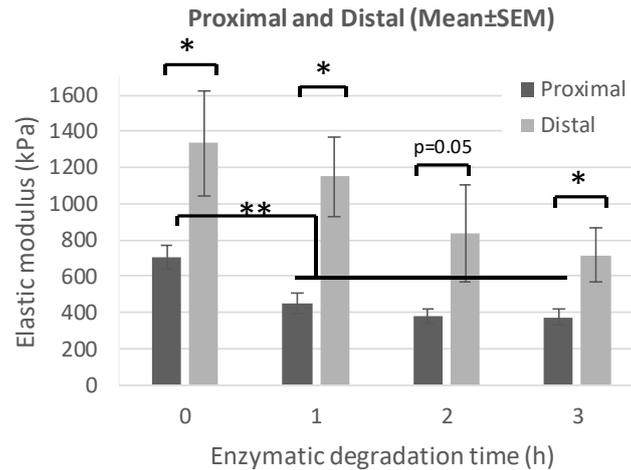

Fig. 9. Statistical analysis of the secant circumferential stiffness in the proximal and distal common carotid segments artery after 0, 1, 2 and 3 hours of collagenase degradation, * : p < 0.05, ** : p<0.01

## 4. Discussion

Enzymatic digestion of the ECM in arteries was instrumental at the early stages of vascular biomechanics [10-12] to demonstrate the respective roles of elastin and collagen in the biomechanical behavior of arteries. However, enzymatic degradation is also a natural effect participating to arterial remodeling, adaptation, and also involved in the development of pathologies such as aneurysms. In this work, we reproduced conditions as close as possible to *in vivo* conditions to investigate the effects of enzymatic degradation on the biomechanical behavior of arteries. For that, arteries were tested in an incubator, and subjected to pulsatile inflation loading while maintained at their *reference* axial stretch. To our best knowledge, this is the first study achieving such conditions in an *in vitro* environment to study enzymatic degradation. This appears as essential to have an insight into proteolytic remodeling as a number of studies have previously shown that the degradation behavior of arterial tissue is strain rate dependent [17], and is also affected by the strain biaxiality [18]. Therefore, tests should reproduce as closely as possible the *in vivo* biaxial loading conditions to characterize proteolytic effects. Pulsatility of the inflation loading is also an important aspect, and we were the first to investigate enzymatic degradation of arteries under pulsatile inflation.

After immersion in the collagenase solution, the fraction of intact fibers decreased significantly in the carotid segments, and this manifested by a decrease of the average secant circumferential stiffness, estimated between 80 mmHg and 140 mmHg. It was important to estimate the effect of enzymatic degradation on stiffness properties under the *in vivo* loading conditions, as this relates the histomechanical effects of enzymatic degradation to the effects on the proper function of arteries. Biomechanical effects may be somehow confusing if we do not consider properly the *in vivo* loading conditions. For instance, tensile tests were carried out by Yuan et al [20] on arteries following enzymatic degradation, and they showed that, at the same mean force, collagen fibers operated at a higher portion of their stress-strain curve, which resulted in an increase in the elastic modulus [27].

In summary, our results show that the main effect of ECM degradation by collagenase is a softening of the biomechanical response, which is a common effect of microstructural damage in materials [28]. Damage has been a topic of intense research in computational mechanics of soft tissues and arteries [29]–[35]. However, authors usually investigated mechanically induced damage in situations of overloading preceding rupture [36]. In aneurysms, damage would rather be induced chemically and occur progressively. A better understanding of the relationship between the microstructural effects related to damage and their macroscopic manifestations is essential for computational analysis of aneurysm development. Therefore, it appears that enzymatic degradation is a very interesting experimental model of investigation.

An important aspect of proteolytic effects and related remodeling are the regional effects. For instance, it was shown that the strength of aortic aneurysm is usually not spatially uniform [37], [38]. A number of studies also showed significant evidence of regional variations of both passive mechanical properties and of elastin, collagen and SMC contents [24], [25], [39]. In the current study, we found statistical significant differences between the mechanical behavior of proximal and distal locations in the circumferential direction of swine carotid arteries, in agreement with the study presented by Garcia et al [25]. Regional variations of material and structural properties can be explained by the effects of vascular adaptation, which try to maintain stresses in the arterial wall at homeostasis [40].

The effect of collagenase on the stiffness did not show a major difference between proximal and distal samples, whereas the histology shows larger degradation in distal samples. But samples on which the collagenase treatment was more pronounced also showed a larger dilation and the main consequence was that these samples had a larger maximum stress after collagenase treatment. So the direct effects of collagenase treatment may indeed be larger on distal samples, but a direct effect may be a larger increase of diameter and then a larger increase of circumferential stress. This subsequent increase of circumferential stress may induce a stiffer response of the tissue, as collagenous tissues become stiffer when they experience larger stresses. So compared at the same stress level, proximal samples would be significantly stiffer than distal ones after collagenase treatment. But compared at the same pressure, the larger loss of stiffness in distal samples is compensated by larger stresses and results in apparent similar stiffness.

Despite our effort in conducting this research with *in vitro* conditions reproducing as much as possible the *in vivo* ones, our tests were carried out on porcine arteries and not human ones. It is still usually accepted that vascular research on swine can be relevant for human due to the large number of similarities between the human and swine cardiovascular systems. We also carried out our tests on carotid arteries, although proteolytic effects seem more relevant for the aorta which is commonly affected by the aneurysmal pathology, whereas carotid arteries are more commonly affected by atherosclerosis. Due to their larger thickness, aortas appeared to be more resistant to the enzymatic degradation process. This prevented us from observing significant biomechanical effects within hours for aortas and this lead us to consider the swine common carotid artery for these tests. Nevertheless, as shown in Fig. 5, the carotid segments underwent radial dilatation under proteolytic effects, as aortas would do when developing an aneurysm. This means that carotid arteries are not immune to proteolytic effects although they do not develop aneurysms *in vivo*.

Another limitation of our study may be related to the chemistry of proteolytic effects. The different types of collagenase produced naturally by vascular cells are collagenases 1, 2 and 3 (MMP1, 8 and 13). Bacterial collagenase was used in our study, permitting fast proteolytic effects but with possible differences of the proteolytic effects naturally occurring in blood vessels. Moreover, in our experiments the pressure cycles were imposed between 30 and 150 mmHg instead of between 80 and 140 mmHg (physiological range). The larger value (150 mmHg) was chosen on purpose. It can reproduce some possible hypertension, but it remains in a decent physiological range as this value can be reached every day in every individual. The smaller value (30 mmHg) is not physiological but had to be imposed due to technical constraints. However, the impact of this is marginal onto the main results of our study, as we look at the mechanical effect (secant stiffness) of the collagenase treatment across the 80 – 140 mmHg range and not across the 30 – 80 mmHg range. We showed that samples experience larger stresses after the collagenase treatment, but a smaller stiffness. Mechanical damage may be induced by the large stresses (above 400-500 kPa, which occurred for only one sample) but not by the ones reached for pressures across the 30 – 80 mmHg range. The fatigue response of a material depends on the magnitude of the stress range, and can be different for a test between 30-140 mmHg or between 80-140mmHg. But our tests only lasted 3 hours, which represents about 10000 cycles. This is not enough to exhibit a sensitivity to fatigue effects (occurring usually after at least $10^6$ cycles). Therefore, damage effects in our experiments mainly result from collagenase treatment only.

Additionally, with the collagenase treatment, the carotid samples dilate (diameter increase), but sometimes the diameter increase is not uniform circumferentially and some samples even developed a "blister-like" dilation. These heterogeneous dilations may disturb diameter measurements, which may lead to non-coherent results on some samples as shown in Figures 3 and 4 (2h degraded sample stiffer than 1 hour degraded sample). Moreover, the deformations of these samples under pressure variations may be rather complex and induce the larger hysteresis.

Another limitation may be the effect of freezing and thawing on the structure and properties of arteries. This has been previously investigated by several authors. Studies that investigated the effect of freezing report rather diverse results. Stemper et al (2007) [20] report no effect of freezing on porcine aortas. Venkatasubramanian et al (2006) [21] report an effect of freezing on the mechanical response in the low-strain region for porcine femoral arteries.

Finally, Chow and Zhang (2011) [22] report no effect in low-strain region but increase in the stiff region for bovine thoracic aortas.

In our study, although it is not possible from literature results to correctly assess the effect of freezing and thawing, all the samples were treated using the same freezing and thawing protocol. This means that the comparative conclusions for instance between distal and proximal carotid arteries, or between untreated and treated arteries, should not be affected by freezing and thawing. Moreover, the circumferential stiffness found for the untreated carotid arteries, for instance of 750 kPa for proximal samples, is in very good agreement with the range order of reported values for fresh porcine carotid arteries [23].

Due to these limitations, it is important to remember that our experiments are not an *in vitro* model of a particular arterial disease, but still, by providing a collection of enzymatically damaged arteries, they may aid in the development of computational models correlating structure and properties of blood vessels. Our experiments, like *in vivo* conditions, induce two opposite effects when collagen fibers are damaged: a structural stiffening at constant pressure due to the induced distension and a material softening due to the material damage. Biomechanical computational models of blood vessels should be tested in these specific conditions for proving their significance for *in vivo* applications.

Future work aims at repeating the same experiments while measuring the whole strain fields of the arteries using digital image correlation [36]. This will permit to have an insight on the local effects of collagenase treatment, whereas the current study was focused on global effects. Moreover, the thickness of the vessel plays an important role on the penetration of collagenase and on the induced proteolysis. The effects of thickness variations across the vessel wall should also be considered in future work.

## 5. Conclusion

In summary, in the current paper, we assessed the histomechanical effects of collagen proteolysis in arteries under loading conditions reproducing *in vivo* conditions. A softening of the arterial wall and a progressive radial dilatation were observed after the beginning of collagenolysis. Eventually, these results constitute an important set of enzymatically damaged arteries that may be useful for computational damage mechanics of arteries.

**Acknowledgements**

The authors are grateful to the European Research Council for ERC grant Biolochanics, grant number 647067.